\documentclass[prb,twocolumn,showpacs,preprintnumbers,amsmath,amssymb,superscriptaddress]{revtex4}
\usepackage{graphicx}
\usepackage{dcolumn}
\usepackage{bm}
\usepackage{color}
\usepackage{subcaption}
\usepackage{xfrac}
\usepackage{epstopdf}
\usepackage{epsf}
\usepackage{natbib}

\begin{document}
\title{Melting curve of elemental zirconium}

\author{Paraskevas Parisiades}\email{paraskevas.parisiadis@upmc.fr}
\affiliation{Institute of Mineralogy, Materials Physics and Cosmochemistry, 4 Place Jussieu, F-75005 Paris, France }
\author{Federico Cova}
\affiliation{ID15B Beamline, European Synchrotron Radiation Facility, 71 Avenue des Martyrs, 38000 Grenoble, France}
\author{Gaston Garbarino}
\affiliation{ID27 Beamline, European Synchrotron Radiation Facility, 71 Avenue des Martyrs, 38000 Grenoble, France}
\vskip 0.7cm

\begin{abstract}
Melting experiments require rapid data acquisition due to instabilities of the molten sample and optical drifting due to the high required laser power. In this work, the melting curve of zirconium has been determined for the first time up to 80 GPa and 4000 K using in-situ fast x-ray diffraction (XRD) in a laser-heated diamond anvil cell (LH-DAC). The main method used for melt detection was the direct observance of liquid diffuse scattering (LDS) in the XRD patterns and it has been proven to be a reliable melting diagnostic. The effectiveness of other melting criteria such as the appearance of temperature plateaus with increasing laser power is also discussed.

\end{abstract}

\maketitle
\bigskip

Zirconium (Zr) and its alloys have a very wide range of applications, from the chemical processing (as corrosion resistant materials) to the semiconductor industry\cite{Durham1986}. Moreover, its good strength and ductility at high temperatures and the low thermal neutron cross-section absorption make it an ideal material for use as cladding at nuclear reactors\cite{Motta2007}. Alloys of Zr with Cu, Al, Ti and Ni have been demonstrated to exhibit extraordinary glass forming ability\cite{Wang2004}, while metallic glass formation in single-element zirconium has also been discovered, with a wide stability in high pressure and temperature conditions\cite{Zhang2004}.
\par
Zirconium is a d-orbital transition metal with a rich and interesting phase diagram. At ambient conditions it crystallizes to an hcp structure ($\alpha$-phase), while at temperatures higher than 1136 K it transforms to a bcc ($\beta$-) phase. By increasing pressure at ambient temperature it transforms to another hexagonal, but not close-packed, called the $\omega$-phase and then back to $\beta$-phase around 35 GPa\cite{Akahama1991,Ono2015, Zhang2005}. Similar transitions also occur in other group IV transition metals, such as Ti and Hf, and it seems that the electronic transfer between the broad sp band and the much narrower d band is the driving force behind those structural transitions.
\par
The high melting point (2128 K) of Zr often classifies it as a refractory metal. Although there are some works in the high temperature behavior of zirconium at high pressures \cite{Jacobsen2017, Zhao2005, Kutsar2006, Zhang2017}, its melting curve has not yet been investigated and this absence of experimental data has strongly motivated this study. On the other hand, the high pressure melting of transition metals has always been a subject of intense debate, because of the large uncertainties in the temperature measurements and the criteria used to identify the melting, so that different approaches can yield very different results. In most cases shock wave (SW) experiments and molecular dynamics (MD) calculations provide dramatically steeper curves than those obtained with the laser speckle method in a LH-DAC, where the melting is visually detected by observing the movements on the sample surface during heating. Tantalum is a good example of such a controversy, with melting temperatures that differ thousands of K at 100 GPa by applying different experimental techniques\cite{Dai2009, Errandonea2001, Errandonea2003, Dewaele2010}. Another more recent example is that of iron\cite{Boehler1993, Anzellini2013}, where the speckle method was found to coincide with the onset of dynamic recrystallization rather than melting. The observation of a temperature plateau versus laser power during a laser heating experiment has also been suggested as a melting diagnostic that works fine in the case of nickel\cite{Lord2014}, but has not been always reliable. Geballe and  Jeanloz \cite{Geballe2012} have argued that the latent heat of the melting inside the LH-DAC is insignificant compared to the heat provided by the lasers, and proposed that the observed temperature plateaus in the laser heating of metals are mainly associated with discontinuous increases in reflectivity rather than melting. Lately, energy dispersive X-ray absorption spectroscopy (XAS) has been also used as a method of detecting melting in metals under high pressure\cite{Torchio2016, Aquilanti2015}, by tracking the disappearance of the shoulder of the XANES signal, as well as the flattening of the first few oscillations. One of the most consistent and recent methods for determining melting is with in-situ synchrotron X-ray diffraction (XRD), by the direct observation of the first liquid diffuse scattering (LDS) signal in the XRD patterns\cite{Lord2014, Anzellini2013, Stutzmann2015}, as the temperature is gradually increased.
\par
In this work we investigate the melting curve of the $\beta$-phase (cubic bcc) of zirconium at pressures up to 80 GPa. To our knowledge there are not any scientific data concerning the melting curve of zirconium. We apply the synchrotron XRD technique in a LH-DAC, which permits to track any chemical reactions in the sample, such as the formation of carbides due to the reaction of the sample with the diamond anvils, and allows for in-situ temperature measurements using pyrometry. We compare the effectiveness of two different melting criteria: the appearance of the first liquid signal in the XRD patterns by increasing temperature, and the observation of temperature plateaus with increasing laser power.
\par

Several membrane-driven diamond anvil cells with culet sizes ranging from 150 to 350 $\mu$m were used to pressurize the samples. Rhenium gaskets, pre-indented to 30 $\mu$m and drilled with a Nd:YAG pulsed laser formed the sample chamber. The sample assembly consisted of high purity (99.2\%) flattened grain Zr pieces of thicknesses around 10 $\mu$m. Having a high purity sample for melting studies is important, since a large amount of impurities could lower the free energy of the system and thus lower the melting point. The sample was embedded between two thin disks of KCl in order to provide thermal and chemical insulation from the diamonds during the laser heating experiment. KCl also serves as a soft pressure transmitting medium and a pressure calibrant. A ruby chip was also installed next to the sample for the estimation of the initial compression of the cell before heating. All the DAC loadings have been carried out in a glove box under an argon atmosphere and the KCl was dried at 100 $^\circ$C for several hours to avoid any amount of water that could trigger chemical reactions between the sample and the diamond anvils\cite{Dewaele2010}.
\par
The in-situ synchrotron x-ray diffraction experiments have been carried out at the ID27 beamline of the ESRF, using a monochromatic beam with a wavelength of 0.3738 \AA. The beam was focused to a spot of 3x2 $\mu$m$^2$. The XRD data were collected by a MAR165 CCD detector calibrated for sample to detector distance with a CeO$_2$ standard. Typical exposure times were in the order of 5 sec and the 2D images were converted to 1D patterns using the Dioptas software\cite{Dioptas}. The diffractograms were fitted by the Le Bail method using the Fullprof software\cite{Carvajal}.
\par
The samples were heated simultaneously from both sides by two continuous YAG fiber lasers (wavelength 1.064 $\mu$m), providing a maximum combined power of 200 W. The YAG laser is very well absorbed by the surface of opaque metallic samples such as Zr\cite{Shen2001}. However, there can be gradients in the temperature inside the bulk of the sample, especially if the area of the melt is small and thus difficult to detect by XRD. Therefore, the experimental procedure has been devised to minimize the temperature gradients in the samples and in the same time provide definitive melting criteria. The alignment of the x-rays, lasers and pyrometry spot were verified after each heating cycle, and the  temperature between the two sides was kept very similar as we gradually increased the laser power. The laser spot size was slightly defocused to 20 $\mu$m diameter (much larger than the x-ray spot) to obtain a more uniform heating of the sample and reduce the temperature gradients due to the Gaussian shape of the TEM$_{00}$ mode of the lasers.
\par
The temperature has been determined by pyrometry measurements using the online system of ID27, with reflective, Schwarzschild objectives which are by construction free of chromatic aberrations. Using this setup, the uncertainty related to the radial temperature gradient is less than 50 K, while the axial component is below 100 K, giving a maximum uncertainty of 150 K\cite{Schultz2007}. The optical path of the collected black body radiation has been calibrated using a tungsten ribbon lamp, with a reference temperature of 2500 K. The temperature is given by the Planck fit in the wavelength window 600-900 nm. Despite the absence of chromatic abberations in the optics, the temperature uncertainty can be much higher, especially at higher temperatures, because of the wavelength dependency of the emissivity. For this reason we have compared the Planck fit with two-color pyrometry measurements, as discussed in \cite{Benedetti2004}. The pressure was determined before and after heating by the fluorescence ruby scale\cite{Dewaele2008}. The final pressure was estimated by the XRD measurements, correcting for thermal pressure by using the thermal equation of state of KCl \cite{Dewaele2012}.
\par
\begin{figure}
\centering
\includegraphics[width=0.9\linewidth]{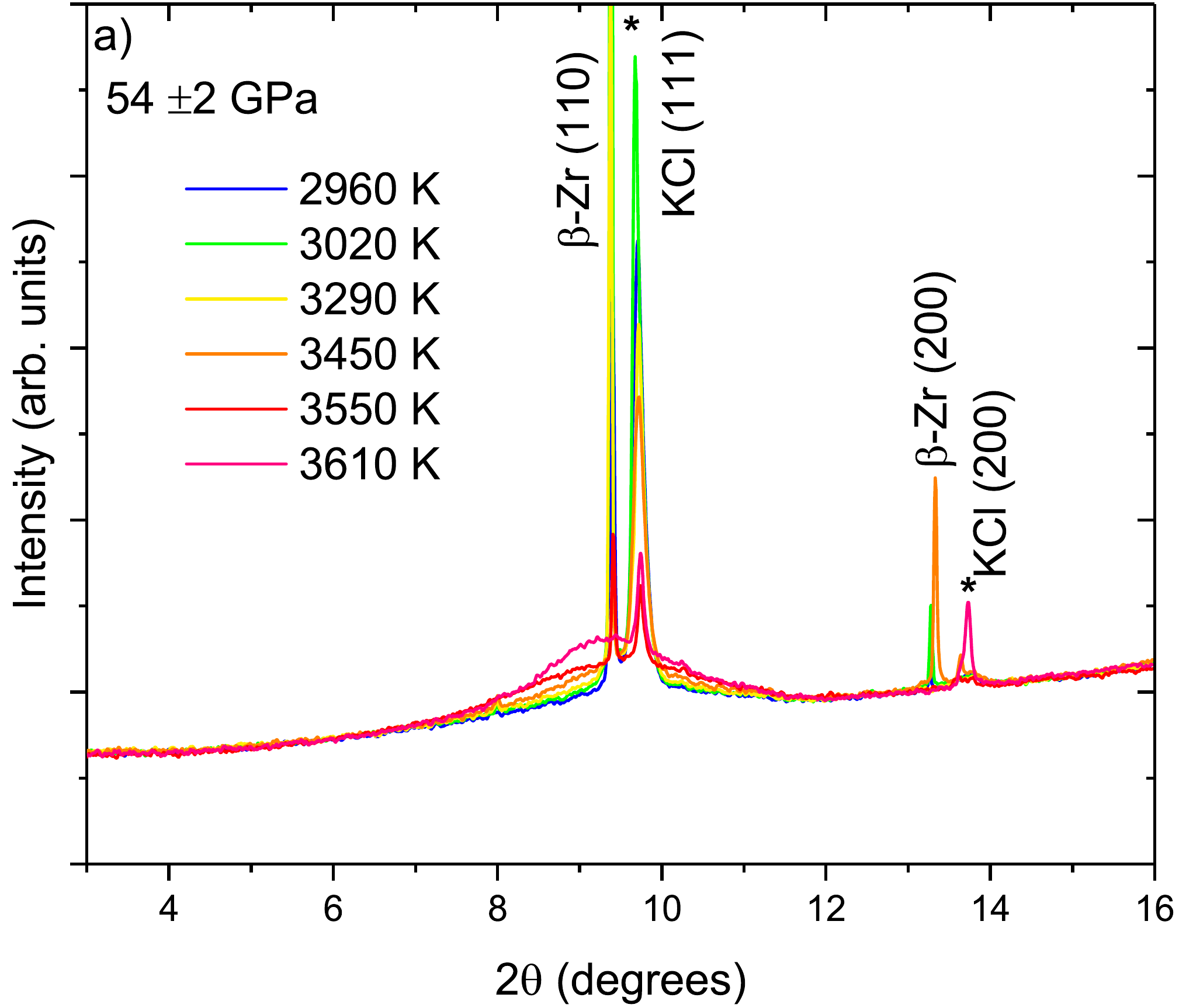}
\includegraphics[width=0.9\linewidth]{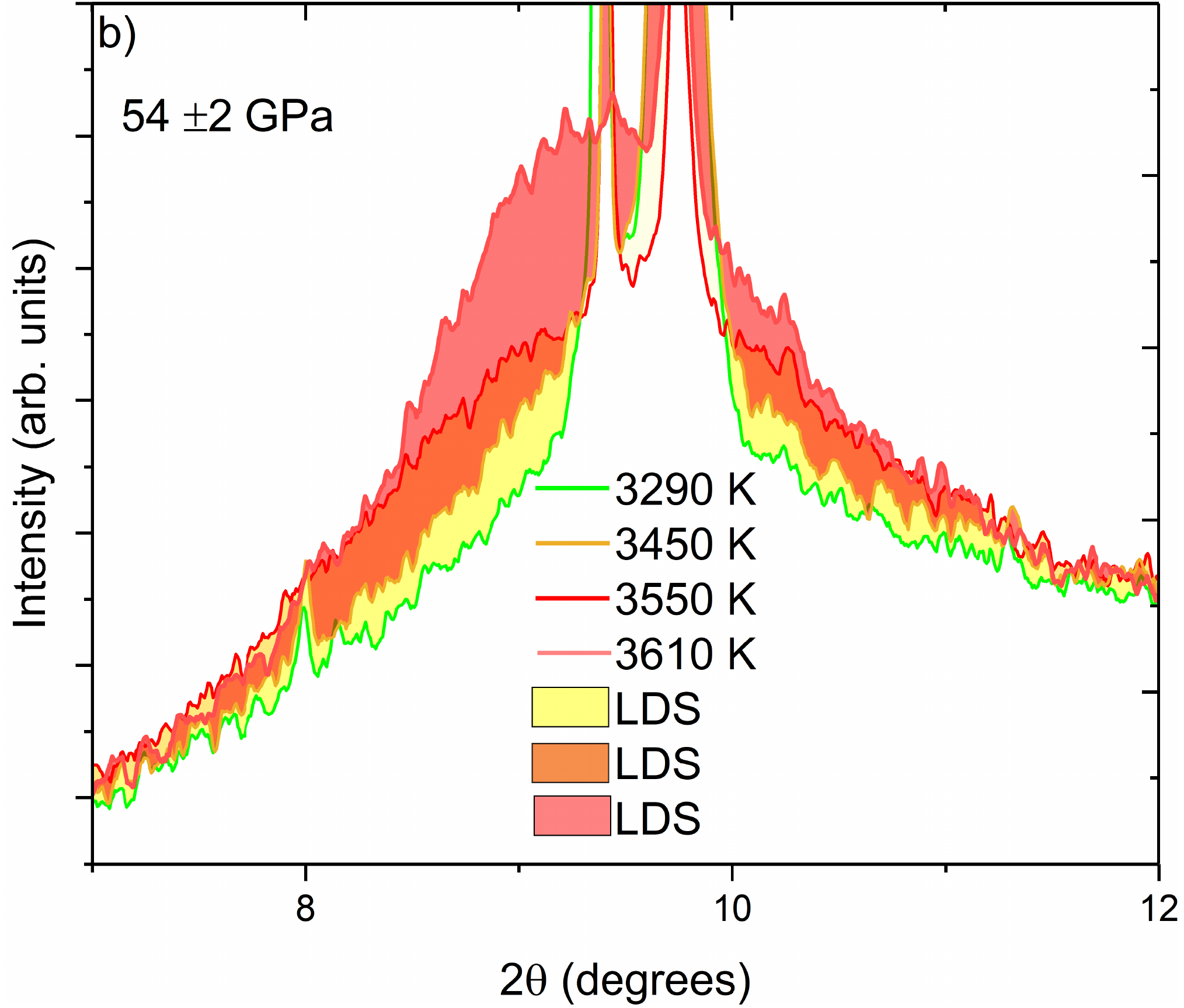}
\includegraphics[width=0.9\linewidth]{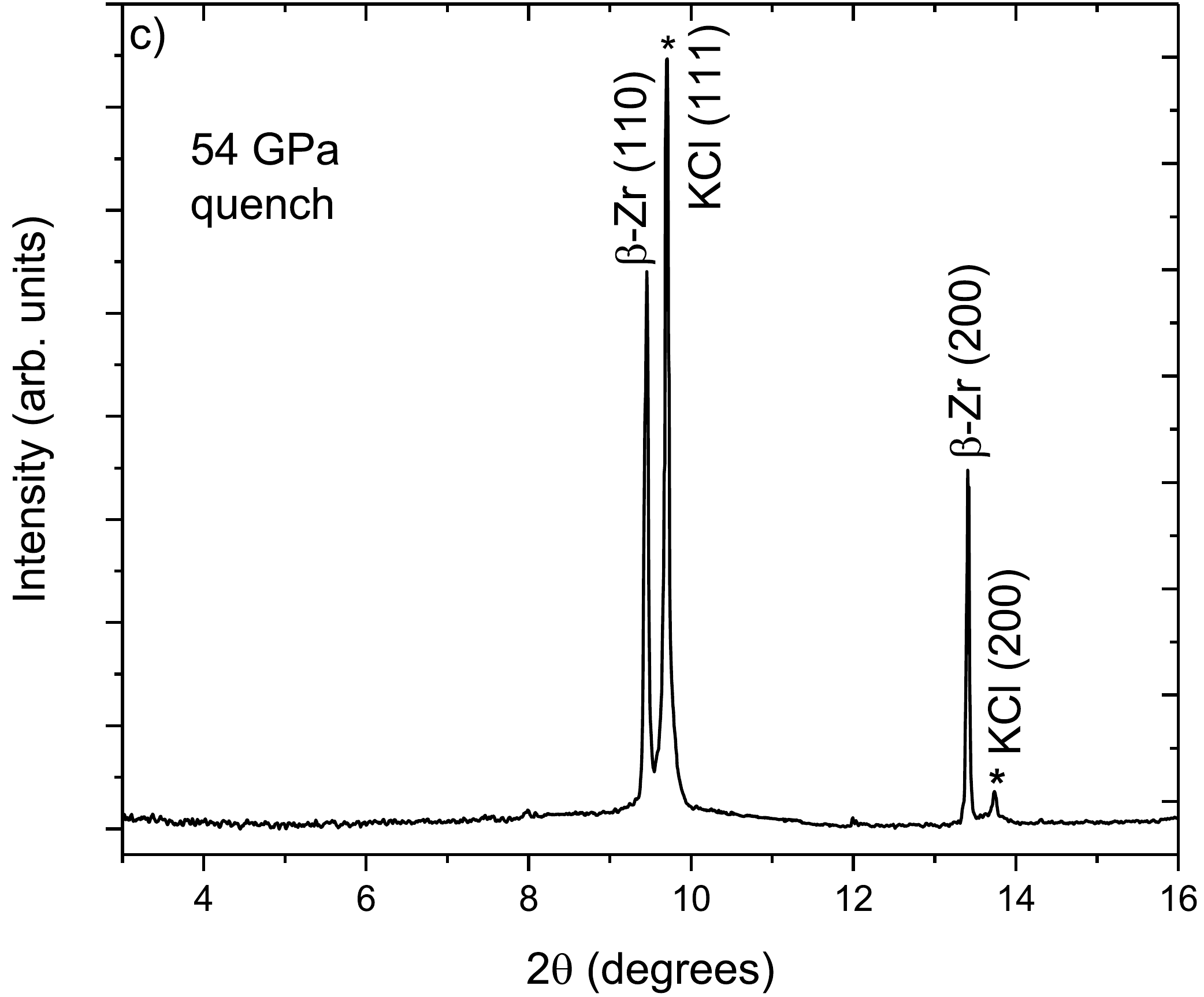}
\captionsetup{justification=raggedright,singlelinecheck=false}
\caption{(Color online) XRD patterns for a heating run at 54 GPa: a) Full spectra, indicating the main peaks of $\beta$-Zr and KCl. b) Zoom of the molten region for different temperatures. c) Quenched data, indicating only Zr and KCl peaks, thus no ZrC was formed upon heating.}
\end{figure}
\par

 At the high temperatures provided by the lasers, Zr was always in the $\beta$-phase. We were able to record the appearance of liquid Zr for six different pressures between $\sim$29 and $\sim$80 GPa. Every pressure point corresponds to a different DAC loading. For most pressure points, we have carried multiple temperature runs by laser heating a different, fresh region of the sample. The main criterion for the detection of melting in our experiments was the first observation of a diffuse liquid signal in the diffraction patterns. The liquid signal is mainly located around the (110) peak of $\beta$-Zr (Figs.1a\&b). This peak is very close to the main (111) reflection of KCl , however the melting curve of KCl is much stiffer than the one of Zr\cite{Boehler1996}, meaning that the observed melt is indeed Zr. Zr can form glasses at high temperatures, especially at high pressures\cite{Zhang2004}, so we were careful to take the melting points in this study outside the glass formation line. As a verification, the diffuse signal disappears in the quenched (Fig.1c) data (i.e. data taken right after switching off the lasers,) meaning that it cannot correspond to an amorphous/glass phase. The quenched pattern of Fig.1c also shows that there was no carbide (ZrC) or oxide (ZrO$_2$) formation during laser heating due to interaction with the diamond anvils or the sample environment. Chemical interactions such as carbide formation can be a big problem in some laser heating experiments, as in the case of Ta.\cite{Dewaele2010}
\par
In Figs.1a\&b we plot the XRD data for a selected pressure of $\sim$54 GPa and increasing temperatures. It can be clearly observed that the liquid signal in XRD increases with increasing temperature, indicating that the amount of molten Zr is also increasing. In the same time the main peak of Zr at $\sim$9.3 degrees decreases with increasing temperature, showing the reduction of solid amount in the sample. However, solid Zr persists for temperatures much higher than the melting point in our XRD patterns, because of the temperature gradients between the surface and the bulk of the sample, which is scanned thoroughly by the x-rays. For a given laser power, the temperature of each side was measured several times before and after the XRD pattern, from both sides, in order to verify that the temperature is not shifting significantly between measurements. The measured x-ray background is constant at low temperatures and starts increasing gradually after a given temperature, providing the signature of melting. The melting temperature was defined as the average temperature between the last solid-only pattern and the first observed pattern with liquid signal as we increased the temperature. For the example of Fig.1a, the liquid signal starts to appear between 3290 and 3450 K and thus the melting temperature was defined to be (3290+3450)/2=3370 K at 54 GPa. In Fig.2 the two-dimensional XRD data are shown for two cases: a) hot but solid $\beta$-Zr (2960 K) and b)solid and liquid mix of Zr well above the melting line (3610 K). The diffuse scattering background is obvious and even at the highest temperature reached in this run (3610 K), some signal from solid Zr persists, although it is reduced significantly, meaning that even at this temperature Zr is only partially melted. KCl remains unmelted, since it has a much higher melting point\cite{Boehler1996}, however it has started growing single crystals, as it can seen from the spots in the diffraction patterns of Fig.2b.

\par
\begin{figure}
\centering
\begin{subfigure}{0.45\linewidth}
\includegraphics[width=\textwidth]{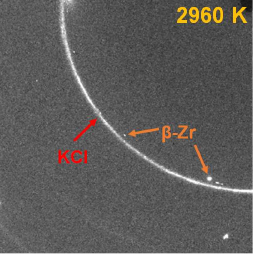}
\end{subfigure}
\hfill
\begin{subfigure}{0.45\linewidth}
\includegraphics[width=\textwidth]{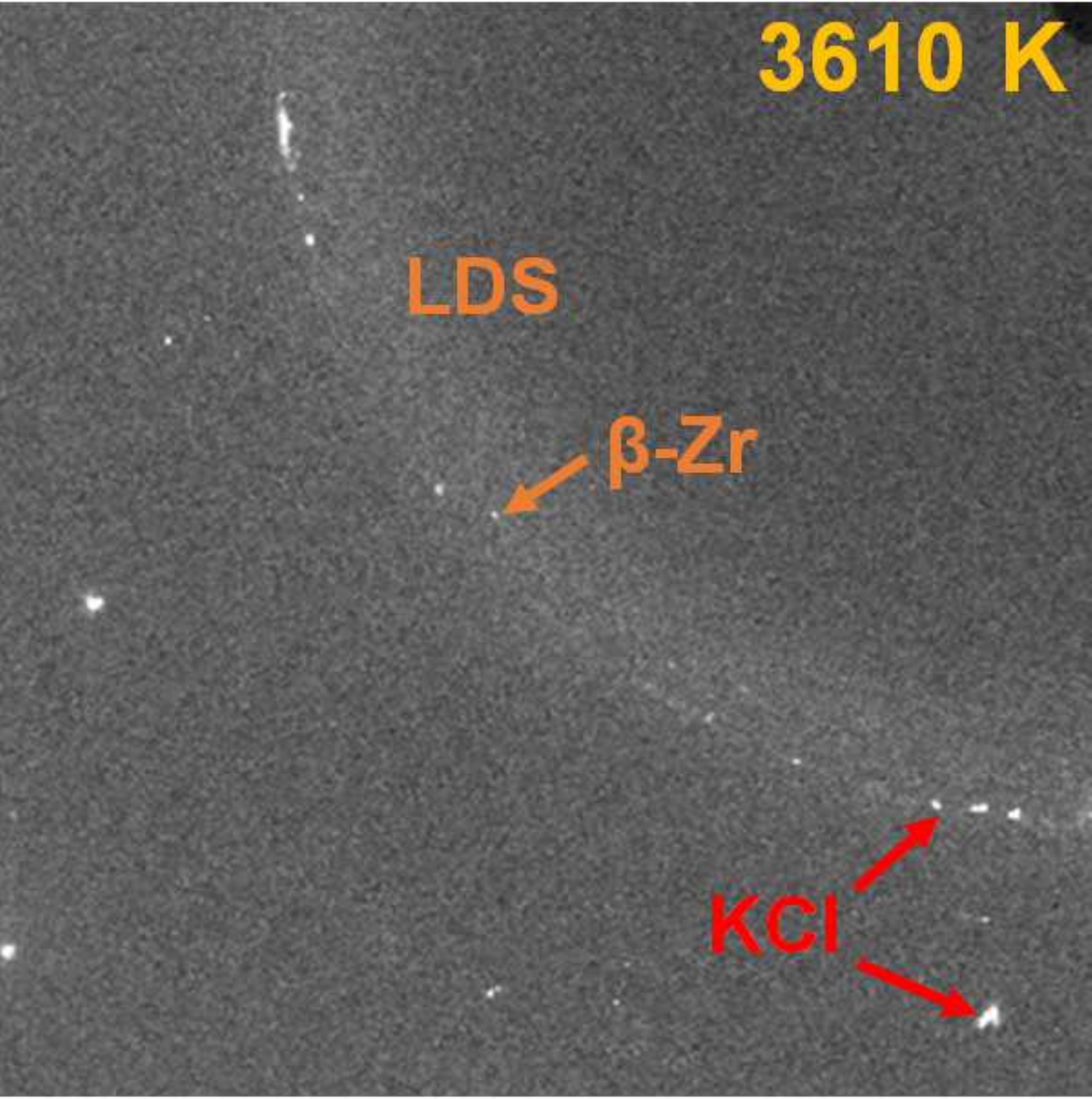}
\end{subfigure}
\captionsetup{justification=raggedright,singlelinecheck=false}
\caption{(Color online) 2D XRD spectra at a thermal pressure of 54 GPa: (a) solid $\beta$-Zr at 2960 K and (b) solid and (mainly) molten Zr at 3610 K.}

\end{figure}

\par
Fig.3 resumes the experimental pressure-temperature conditions for all the XRD patterns recorded.  Zr crystallizes upon heating and these small crystals appear and disappear in the 2D diffraction patterns with increasing temperature due to small movements in the surface of the sample. This effect has been referred to as “fast recrystallization” and in many cases it can lead to an underestimation of the melting point. In some previous works using optical-based diagnostics, such as the speckle method,  the melting was attributed to this "fast recrystallization" and yielded temperatures much lower than the actual melting temperature, especially at higher pressures\cite{Anzellini2013,Dewaele2010}. In our data the threshold of fast recrystallization and melting seem to differ around 1000 K for most pressure points. In the absence of other melting data for Zr, comparing with a recent work on Ti, another d-orbital transition metal, has shown discrepancies of 500 K between fast recrystallization and melting\cite{Stutzmann2015}, while in Fe the difference was found to be close to 1000 K\cite{Anzellini2013}.
\par
Hrubiak et al.\cite{Hrubiak2017}, on their recent work on Mo, they discussed thoroughly the phenomenon of fast recrystallization by observing the quenched data obtained from different temperatures. They observed the appearance of preferred orientations on the quenched data obtained from a recrystallization temperature (i.e. before melting). This preferred orientation gave place to a fine grain structure (with random orientation) for quenches obtained from temperatures above melting. In our case we have followed a different methodology since we quenched the sample after obtaining evidence of melting from the in-situ diffraction patterns, so it is not possible to define melting by observing the microstructure changes in Zr upon heating. However, we have observed a fine-grain structure for Zr in the quenched patterns obtained after melting (as shown in the Supplemental Material), in good agreement with the work of Hrubiak et al.\cite{Hrubiak2017}.
\par
\begin{figure}
\centering
\includegraphics[width=0.9\linewidth]{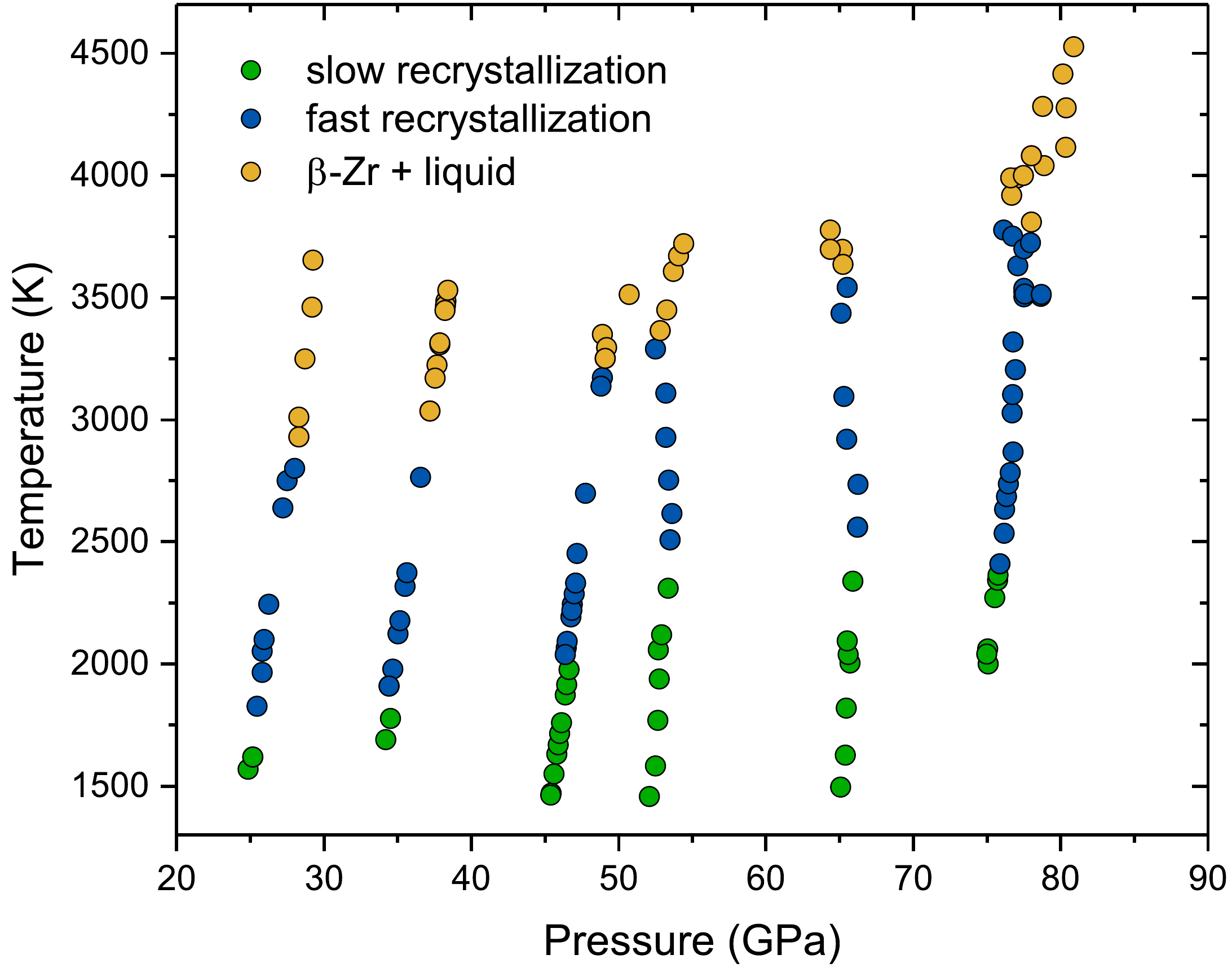}
\captionsetup{justification=raggedright,singlelinecheck=false}
\caption{(Color online) Experimental points for LH-DAC $\beta$-Zr and KCl.}
\end{figure}
\par
Another method that has been thoroughly used for the estimation of the melting line in various works is the presence of temperature plateaus upon heating. Figs.3a\&b show the sample temperature as a function of time, which accounts for increasing laser output. In most of the cases the temperature is increasing rapidly and linearly at low laser power inputs (Fig.3a). In some of the runs the temperature exhibited a plateau, or showed minor fluctuations within error bars near the melting point (Fig3a). However, there are runs where plateaus can be seen well below the detected melting points, that are not associated with fast recrystallization either (Fig.3b). For some experiments the temperature slope during the laser power increase changed several times during a heating cycle (Fig.3b). It is clear that the temperature saturation method as a melting criterion has not a great reproducibility in the case of Zr. There are many reasons why the rate the temperature increases can change in a LH-DAC. The main reason that has been proposed are the discontinuous reflectivity increases in the sample, that cannot however be a reliable indicator of melting since it is not an intrinsic property of materials \cite{Geballe2012}. The increase in the conductivity of KCl with the increasing heat provided by the lasers could also explain why the temperature does not always increase with laser power. Alternatively, the fact that the two criteria of XRD diffuse scattering and temperature plateaus do not always agree in the case of Zr could also be related to  the small differences in the thickness of the sample and KCl layers between the runs, which affect the thermal isolation of the sample, or the heating efficiency inside the DAC. Thus the observation of temperature plateaus for defining melting, although it can work in some cases\cite{Lord2014}, does not seem to be a consistent method because it depends on many parameters that cannot be calculated quantitatively and in-situ in a laser heating experiment. The temperature vs laser power data for runs at different pressures are shown in the supplemental material (SM).
\par

\begin{figure}
\centering
\includegraphics[width=0.9\linewidth]{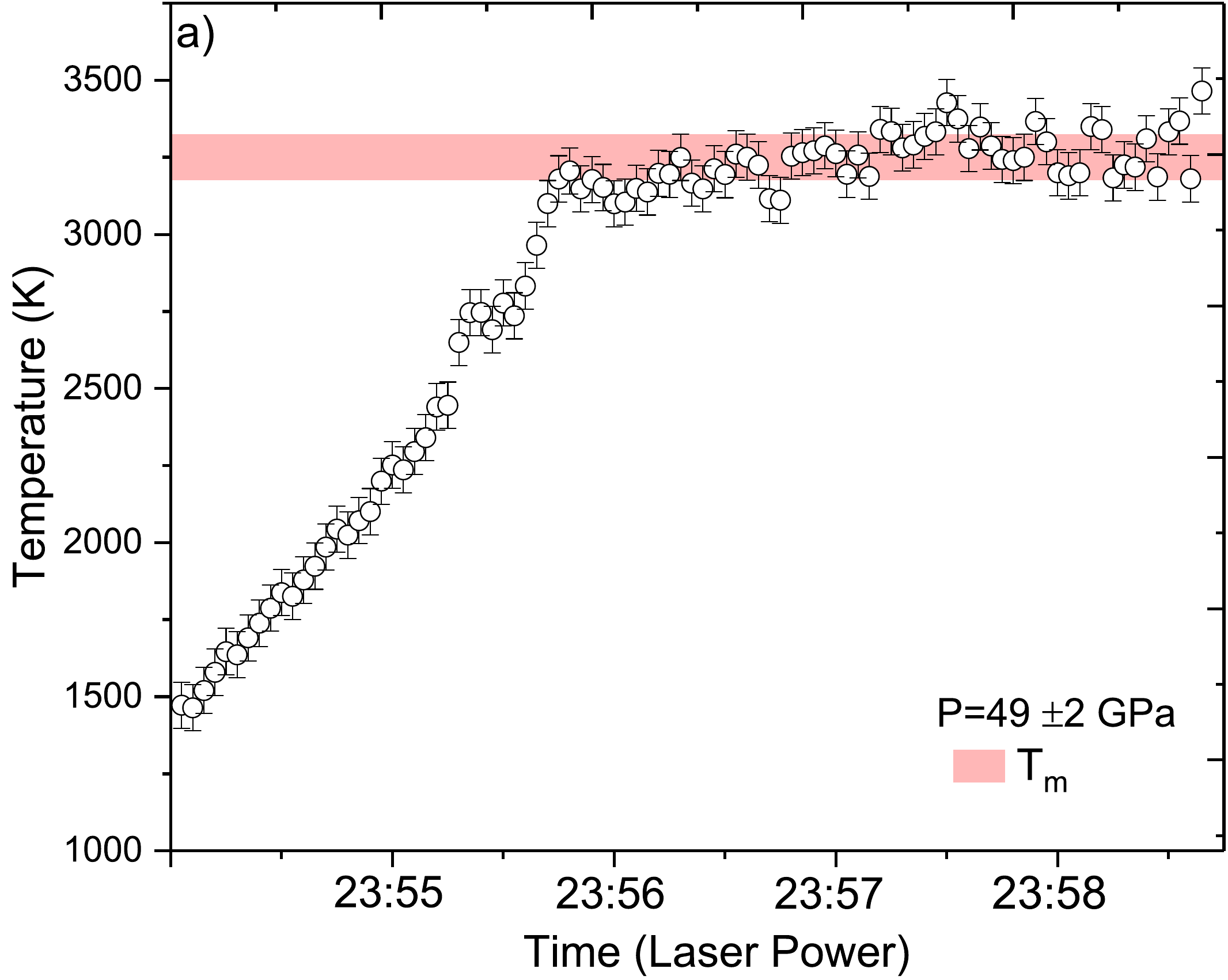}
\includegraphics[width=0.9\linewidth]{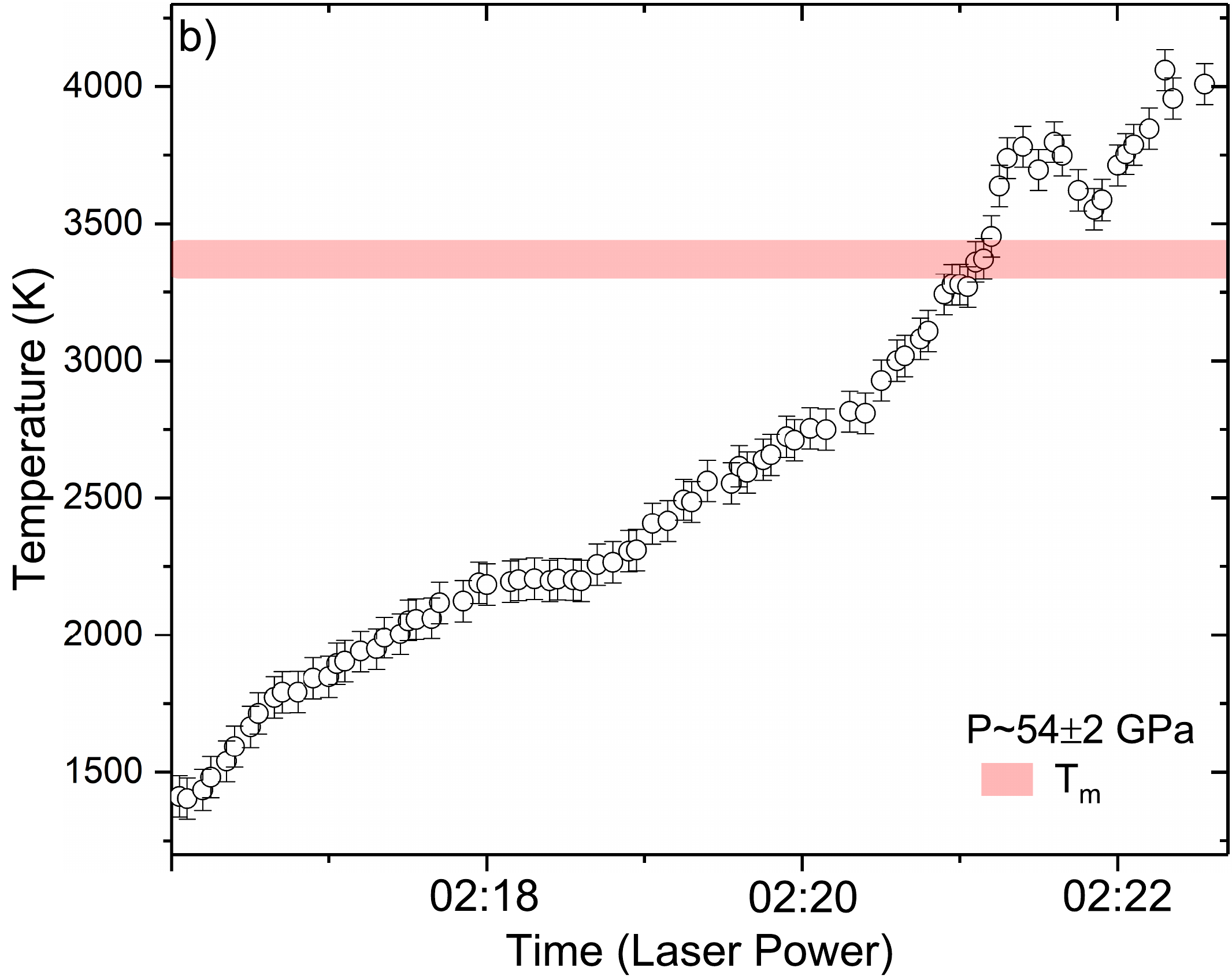}
\captionsetup{justification=raggedright,singlelinecheck=false}
\caption{(Color online) Temperature measurement vs time (laser power) for a heating run at a)65 GPa and b) 54 GPa. The melting temperature defined from XRD (including errors) is denoted by the pink lines.}
\end{figure}
\par

All the melting points obtained in this study are displayed in Fig.5. The melting curve presented is corrected for thermal pressure and is fitted using the Simon-Glatzel equation \cite{Simon1929}, yielding:
\par
\begin{equation}
T_m=T_0\times(\frac{P_m}{42.58\pm 11.34}+1)^{0.56\pm0.09}
\end{equation}

where T$_0$=2128 K is the ambient pressure melting point of Zr, T$_m$ the melting temperature and P$_m$ the melting pressure at each point. The Simon-Glatzel equation gives satisfactory results for the given pressure range, as is the case for many metals\cite{Lord2014, Dewaele2007}.

\par
The melting curve can be also calculated using the Lindemann law \cite{Anderson2000}:
\par
\begin{equation}
T_m=T_0\times(V/V_0)^{2/3}exp(2\gamma_0/q(1-(V/V_0)^q))
\end{equation}

In this equation $\gamma$$_0$=1.01 is the Gruneisen parameter for Zr taken from Goldak et al\cite{Goldak1966} and its volume dependence is taken as q=1 \cite{Boehler2000}. As seen in Fig.5, the results obtained from the Lindemann equation are generally in good agreement with the Simon-Glatzel fit for most of our data. However, it appears that the melting slope obtained this way is slightly higher. Errandonea \cite{Errandonea2005} in a work concerning Mo, Ta and W has found that the Lindemann law could overestimate the melting point of  bcc transition  metals  at  high pressures. This can be due to the fact that the Lindemann law takes into account only the thermodynamic parameters of the solid phase and completely neglects the liquid, which could lead to inaccuracies.

\par
\begin{figure}
\centering
\includegraphics[width=0.9\linewidth]{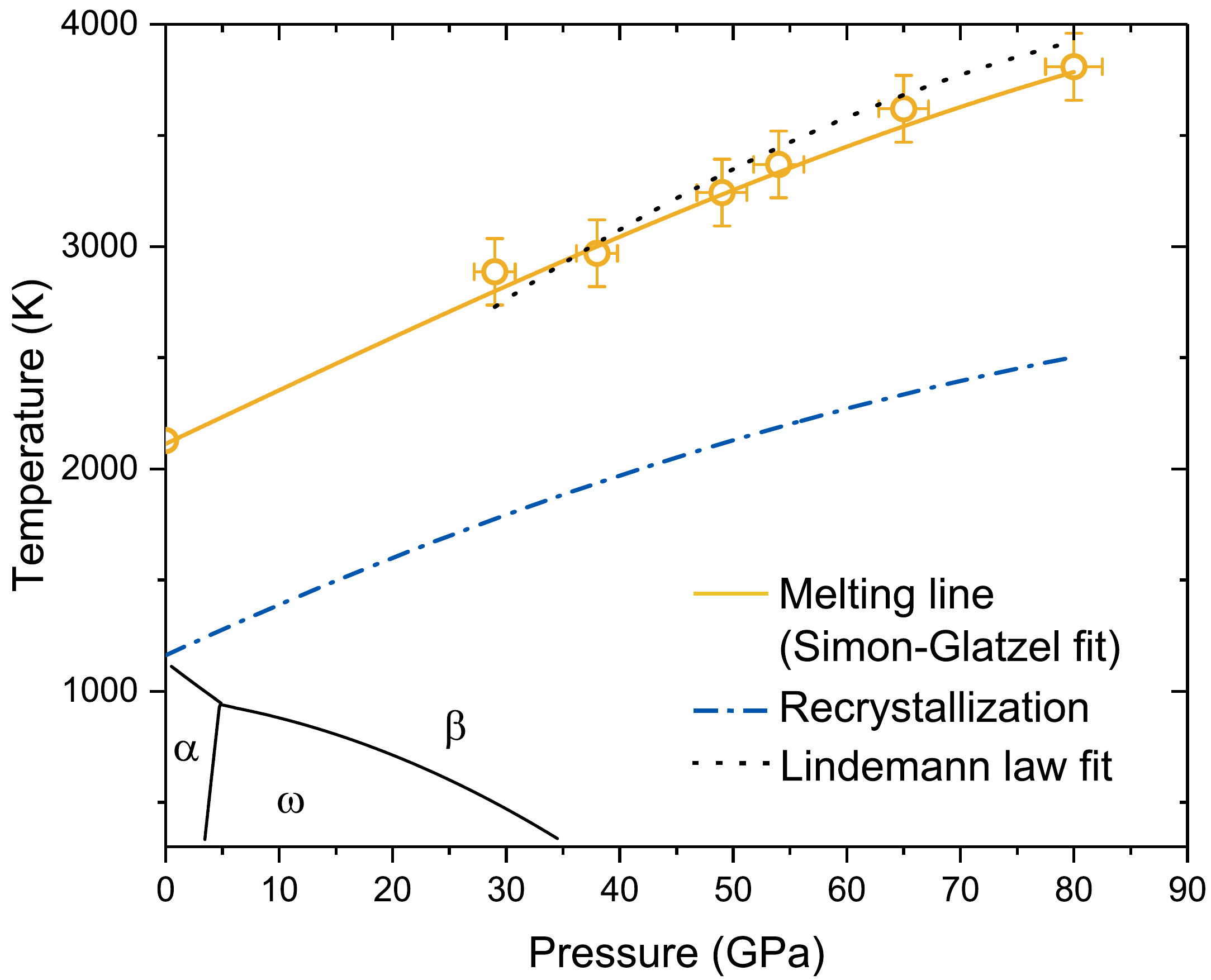}
\captionsetup{justification=raggedright,singlelinecheck=false}
\caption{(Color online) Melting curve of Zr. The yellow solid line corresponds to the Simon-Glatzel curve fit. The fast recrystallization curve is also shown with a blue dashed-dotted line, as derived from the data of Fig.3. The dotted black line corresponds to the melting curve obtained using the Lindemann law.}
\end{figure}
\par
In absence of other experimental data for Zr, we compare the melting curve with the one of Ti from ref.\cite{Stutzmann2015}, which is a transition metal with the same electronic configuration. The melting curves in both Zr and Ti have been determined using similar experimental methods in these two works. By the Simon-Glatzel fit it is possible to calculate a melting slope dT$_m$/dP = 28 K/GPa for Zr, defining it as a low melting slope material. The low slope for Zr (and also Ti) is because of the partially filled d- electron bands and it is a frequent phenomenon in transition metals. It has been proposed that the loss of long-range order due to melting can induce a change in the liquid density of states (DOS) and therefore a decrease in the melting slope, and materials with filled bands and DOS that do not change significantly upon melting, such as Cu, Al, or noble gases, have systematically much stepper melting curves than the partially filled bcc transition metals \cite{Ross2004, Errandonea2010, Errandonea2013}. Also in metals with partial filled d-bands, icosahedral short-range order can be favored energetically in the supercooled liquids and melts, and it has already been observed for Zr\cite{Jakse2003}. These short-range structures can act as impurities that lower the free energy and thus the melting slope\cite{Ross20071, Ross20072}. Although we have only observed partial melt in our data and thus it is not possible to perform a full pair distribution function analysis, we cannot rule out the possibility of an icosahedral short-range order in Zr. In both Zr and Ti, the pressure generates an s-d electron transfer and an increase in the concentration of local structures which  lowers even further the melting slope.

\par

In conclusion, the high-pressure melting curve of $\beta$-Zr has been studied using synchrotron x-ray diffraction in a laser-heated diamond anvil cell. The appearance of liquid signal in the in-situ x-ray patters has been proven to be a reliable diagnostic, that has in the same time the advantage of detecting any possible unwanted reactions in the sample during heating. On the other hand, the observation of temperature plateaus cannot reproduce a specific pattern for melting, and therefore this method was proven unreliable in the case of Zr. This is, to our knowledge the first set of experimental data on the melting curve of Zr and we expect that it will greatly motivate further experimental and theoretical studies on the subject, using different techniques.
\par
\vspace{5mm}
ACKNOWLEDGEMENTS
\par
We acknowledge the European Synchrotron Radiation Facility and especially the beamline ID27, for the beamtime allocated to Proposal No.HC-2799. We would like to thank Guillaume Morard for the fruitful discussions.
\par


\begin{thebibliography}{999}
\bibliographystyle{unsrtnat}
\bibitem{Durham1986} Y. Durham, \textit{Industrial applications of titanium and zirconium}, ASTM, Philadelphia, 1986.
\bibitem{Motta2007} A. T. Motta, A. Yilmazbayhan, M. J. Gomes da Silva, R. J. Comstock, G. S. Was, J. T. Busby, E. Gartner, Q. Peng. Y. H. Jeong, and J. Y. Park, \textit{J. Nucl. Mater.} \textbf{371}, 61 (2007).
\bibitem{Wang2004} W. H. Wang, C. Dong, and C. H. Shek, \textit{Mater. Sci. Eng. R Rep.} \textbf{R 44}, 45 (2004).
 \bibitem{Zhang2004} J. Zhang and Y. Zhao, \textit{Nature} \textbf{430}, 332 (2004).
 \bibitem{Akahama1991} Y. Akahama, M. Kobayashi, and H. Kawamura, \textit{J. Phys. Soc. Jpn.} \textbf{60}, 3211 (1991).
 \bibitem{Ono2015} S. Ono and T. Kikegawa,\textit{J. Solid State Chem.} \textbf{225}, 110 (2015).
\bibitem{Zhang2005} J. Zhang, Y. Zhao, C. Pantea, J. Qian, L. L. Daemen, P. A. Rigg, R. S. Hixson, C. W. Greeff, G. T. Gray III, Y. Yang, L. Wang, Y. Wang, and T. Uchida, \textit{J. Phys. Chem. Solids} \textbf{66}, 1213 (2005).
\bibitem{Jacobsen2017} M. K. Jacobsen, N. Velisavljevic, Y. Kono, C. Park, and C. Kenney-Benson, \textit{Phys. Rev. B} \textbf{95}, 134101 (2017).
\bibitem{Zhao2005} Y. Zhao, J. Zhang, C. Pantea, J. Qian, L. L. Daemen, P. A. Rigg, R. S. Hixson, G. T. Gray, Y. Yang, L. Wang, Y. Wang, and T. Uchida, \textit{Phys. Rev. B} \textbf{71}, 184119 (2005).
\bibitem{Kutsar2006} A. R. Kutsar, I. V. Lyasotski, Am. M. Podurets, and A. F. Sanches-Bolinches, \textit{High Press. Res.} \textbf{4}, 475 (2006).
\bibitem{Zhang2017} H. Zhang, Y. Mo, Z. Tian, R. Liu, L. Zhouc, and Z. Hou, \textit{Chem. Chem. Phys.} \textbf{19}, 12310 (2017).
 \bibitem{Dai2009} C. Dai, J. Hu, and H. Tan, \textit{J. Appl. Phys.} \textbf{106}, 043519 (2009).
 \bibitem{Errandonea2001} D. Errandonea, B. Schwager, R. Ditz, C. Gessmann, R. Boehler, and M. Ross, \textit{Phys. Rev. B} \textbf{63}, 132104 (2001).
 \bibitem{Errandonea2003} D. Errandonea, M. Somayazulu, D.H\"ausermann, and H.-K. Mao, \textit{J. Phys. Condens. Matter} \textbf{15}, 7635 (2003).
\bibitem{Dewaele2010} A. Dewaele, M. Mezouar, N. Guignot, and P. Loubeyre \textit{Phys. Rev. Lett.} \textbf{104}, 255701
    (2010).
\bibitem{Boehler1993} R. Boehler, \textit{Nature} \textbf{363}, 534 (1993).
\bibitem{Anzellini2013} S. Anzellini, A. Dewaele, M. Mezouar, P. Loubeyre, and G. Morard, \textit{Science} \textbf{340}, 464 (2013).
\bibitem{Geballe2012}  Z. M. Geballe and R. Jeanloz, \textit{J. Appl. Phys.} \textbf{111}, 123518 (2012).
\bibitem{Lord2014} O. T. Lord, I. G. Wood, D. P. Dobson, L. Vocaldo, W. Wang, A. R. Thomson, E. Wann, G. Morard, M. Mezouar and M. J. Walter, \textit{Earth Planet. Sci. Lett.}  \textbf{408}, 226 (2014).
\bibitem{Hrubiak2017} R. Hrubiak, Y. Meng, and G. Shen, \textit{Nat. Commun.} \textbf{8}, 14562 (2017).
\bibitem{Torchio2016} R. Torchio, S. Boccato, V. Cerantola, G. Morard, T. Irifune, and I. Kantor, \textit{High Pressure Res.} \textbf{36}, 293 (2016).
\bibitem{Aquilanti2015} G. Aquilanti, A. Trapananti, A. Karandikar, I. Kantor, I. Marini, O. Mathon, S. Pascarelli, and R. Boehler, \textit{Proc. Natl. Acad. Sci.} \textbf{112}, 12042 (2015).
\bibitem{Stutzmann2015} V. Stutzmann, A. Dewaele, J. Bouchet, F. Bottin, and M. Mezouar, \textit{Phys. Rev. B} \textbf{92}, 224110 (2015).
\bibitem{Dioptas} C. Prescher and V. B. Prakapenka, \textit{High Press. Res.} \textbf{35:3}, 223 (2015).
\bibitem{Carvajal} J. Rodriguez-Carvajal, \textit{Physica B} \textbf{192}, 55 (1993).
\bibitem{Shen2001} G. Shen, M. L. Rivers, Y. Wang, and S. R. Sutton, \textit{Rev. Sci. Instrum.} \textbf{72}, 1273 (2001).
\bibitem{Schultz2007} E. Schultz, M. Mezouar, W. Crichton, S. Bauchau, G. Blattmann, D. Andrault, G. Fiquet, R. Boehler, N. Rambert, B. Sitaud, and P. Loubeyre, \textit{High Press. Res.}, \textbf{25}, 71 (2007).
\bibitem{Benedetti2004} L. R. Benedetti and P. Loubeyre, \textit{High Pressure Res.} \textbf{24}, 423 (2004).
\bibitem{Dewaele2008} A. Dewaele, M. Torrent, P. Loubeyre, and M.Mezouar,\textit{Phys. Rev. B} \textbf{78}, 104102 (2008)
\bibitem{Dewaele2012} A. Dewaele, A. B. Belonoshko, G. Garbarino, F. Occelli, P. Bouvier, M. Hanfland, and M. Mezouar, \textit{Phys. Rev. B} \textbf{85}, 214105 (2012).
\bibitem{Boehler1996} R. Boehler, M. Ross, and D. B. Boercker, \textit{Phys. Rev. B}, \textbf{53}, 556 (1996).
\bibitem{Simon1929} F. E. Simon and G. Glatzel, \textit{Z. Anorg. Chem.} \textbf{178}, 309 (1929).
\bibitem{Dewaele2007} A. Dewaele, M. Mezouar, N. Guignot, and P. Loubeyre, \textit{Phys. Rev. B} \textbf{76}, 144106 (2007).
\bibitem{Anderson2000} O. L. Anderson and D. G. Isaac, \textit {Am. Min.} \textbf{85}, 376 (2000).
\bibitem{Goldak1966} J. Goldak, L. T. Lloyd, and C. S. Barrett, \textit{Phys. Rev.} \textbf{144}, 478 (1966).
\bibitem{Boehler2000} R. Boehler, \textit{Rev. Geophys.} \textbf{38}, 221 (2000).
\bibitem{Errandonea2005} D. Errandonea, \textit{Physica B} \textbf{357}, 356 (2005).
\bibitem{Ross2004} M. Ross, L. H. Yang, and R. Boehler, \textit{Phys. Rev. B} \textbf{70}, 184112 (2004).
\bibitem{Errandonea2010} D. Errandonea, \textit{J. Appl. Phys.} \textbf{108}, 033517 (2010).
\bibitem{Errandonea2013} D. Errandonea, \textit{Phys. Rev. B} \textbf{87}, 054108 (2013).
\bibitem{Jakse2003} N. Jakse and A. Pasturel, \textit{Phys. Rev. Lett.} \textbf{91}, 195501 (2003).
\bibitem{Ross20071} M. Ross, R. Boehler, and D. Errandonea, \textit{Phys. Rev. B} \textbf{76}, 184117 (2007).
\bibitem{Ross20072} M. Ross, D. Errandonea, and R. Boehler, \textit{Phys. Rev. B} \textbf{76}, 184118 (2007).
\end{thebibliography}
\end{document}